\documentclass[a4paper,11pt]{article}
\usepackage{jinstpub} % for details on the use of the package, please
\usepackage{balance}
\usepackage{ragged2e}
\usepackage{lineno}
\renewcommand{\justify}{\leftskip=0pt \rightskip=0pt plus 0cm}

\title{\boldmath Study on SiPM performance at low temperatures between $-60^{\circ}$C and $-20^{\circ}$C}

\author[a]{C.~Zhong,\note{Corresponding author.}}
\author[1,a]{F.~J.~Luo,}
\author[a]{B.~Zheng,}
\author[a]{X.~D.~Wang,}
\author[a]{M.~Y.~Bu,}
\author[a]{J.~Zou,}
\author[a]{and M.~N.~Deng}

\affiliation[a]{School of Nuclear Science and Technology, University of South China, \\Hengyang 421001, People's Republic of China}
\emailAdd{luofengjiao@usc.edu.cn}

\abstract{Radon is the main background source of dark matter and neutrino experiments. 
Radon concentration ($\rm mBq/m^3$) measurement by liquid scintillation detector is a highly sensitive method at low temperatures using silicon photomultipliers (SiPMs) arrays. 
The SiPM performance characteristics are closely related to the lower detection limit of the detector. In this study, we built an automatic and accurate low-temperature measurement system to study the single photoelectron (SPE) spectrum, SPE resolution, optical crosstalk, and after-pulse of the SiPM at different temperatures. As a result, we obtained the variation trend of the SiPM parameters at different temperatures, and the SiPM optimal working conditions were obtained, which can improve the detector's sensitivity. 
}


\proceeding{N$^{\text{th}}$ Workshop on X\\
  when\\
  where}

\begin{document} 
\maketitle 
\flushbottom 

\section{Introduction}
Radon is the main background source for many low-background detector experiments~\cite{r1,r2,r3,r4}. Currently, the radon measurement methods worldwide mainly include the ionization chamber, electrostatic adsorption, and liquid scintillator measurement methods. Among them, due to the advantages of a high solubility coefficient and a low background, the liquid scintillator measurement method is widely used in dark matter and neutrino experiments, such as Jiangmen Underground Neutrino Observatory (JUNO)~\cite{juno,juno1} and China Jinping Underground Laboratory (CJPL)~\cite{ud,ud1}. Moreover, they all require high sensitivity for low-background detectors for radon measurements. In liquid scintillators, radon and its daughters generate photons through the $\beta-\alpha$ cascaded decay~\cite{alpha,alpha1}, and photon counts are detected using photomultiplier tubes. However, silicon photomultipliers (SiPMs) are the most popular photon detection devices in the composition of liquid scintillation detectors~\cite{sipm}. As novel low-light-sensing optoelectronic conversion devices, SiPMs have attracted significant attention due to their small size, high gain, fast response, and low cost. It comprises a high-density diode~\cite{pmt} matrix with a common output load, each in a finite Geiger-Muller mode for high gain~\cite{ger,gr,f1}. 

Generally, the detection limit of the detector characterizes the sensitivity in 
radioactivity measurements~\cite{dl}, which refers 
to the minimum expected value of radioactivity that can be detected by a certain measurement method under a certain confidence level. Based on the measurement of radon and its daughters' concentrations, the detection limit is defined as the minimum detectable concentration (MDC)~\cite{mdc}, and is presented in Equation~\ref{a}. 
\begin{equation} \label{a}
MDC=\frac{4.65\sqrt{n_b/t_b}}{\epsilon\cdot\eta\cdot V\cdot\zeta}, 
\end{equation} 
where $n_b$ is the background count rate (cps), $t_b$ is the counting time of the sample and the background (s), $\epsilon$ is the detection efficiency, $\eta$ is the total recovery efficiency during the separation and concentration process, and $V$ is the sample amount ($m^3$), $\zeta$ is the branching ratio of gamma rays or other particles. Therefore, effectively reducing the background count is the key to achieving high sensitivity of the detector, $i.e.$, to reduce the noise influence when the SiPM detects the photon signal in the experiment. We know that the SiPM temperature characteristics make it easy to generate a lot of thermal noise due to the excessive operating voltage~\cite{sipm1,tem1,tem4,tem5}. So, the temperature can seriously affect the background count in detection experiments~\cite{tem,tem3}. Moreover, SiPMs have excellent SPE resolution, and very low noise characteristics at low temperatures, which can effectively reduce the detection limit of the detector and improve the detection sensitivity of liquid scintillation detectors to the decay concentration of radon and its decay daughters. 

Moreover, the sensitivity of the detector is also related to the light yield of the liquid scintillator. In low-temperature experimental environments, the light yield of the liquid scintillator increases by~$2$\% when the temperature decreases by $10^{\circ}$C from room temperature~\cite{yield}. Therefore, by reducing the experimental temperature reasonably, the light yield can be effectively be increased, thereby improving the detector's sensitivity.

In this study, we design the low-temperature measurement system to systematically study the SPE~\cite{fzw,tem5} spectrum, gain, optical crosstalk, after-pulse, time resolution, and SPE resolution of the SiPM under different temperatures. By analyzing the amplitude and charge spectrum of the pulse signal, we can obtain the low-temperature effects on the performance characteristics of SiPMs. Moreover, the optimal working conditions of SiPMs are obtained with lower noise and better SPE resolution.

\section{Experimental system}

\subsection{Measurement system}
\begin{figure}[htbp] \centering
	\setlength{\abovecaptionskip}{-1pt}
	\setlength{\belowcaptionskip}{10pt}
	\includegraphics[width=12.0cm]{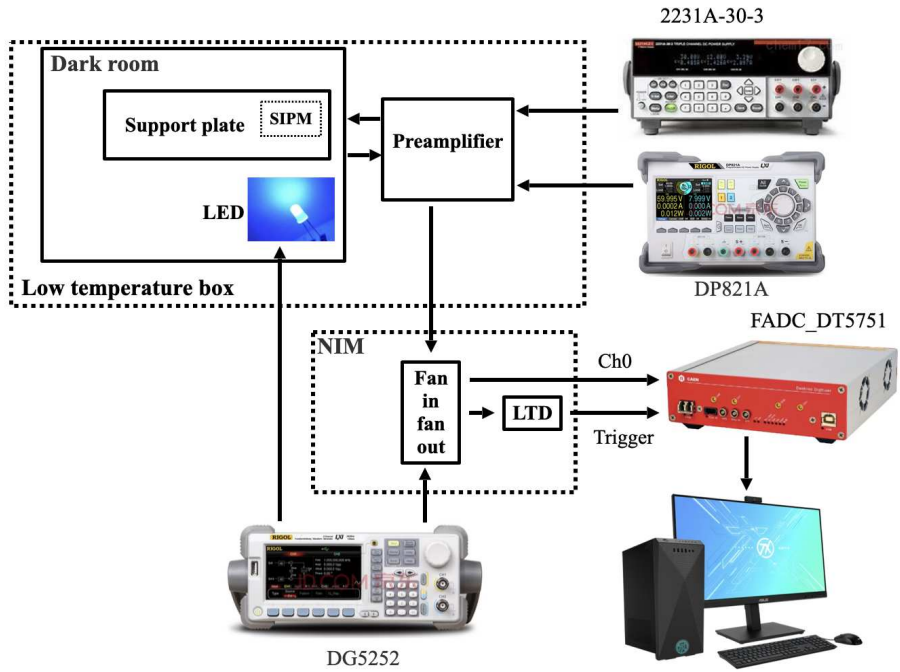}
	\caption{The low temperature measurement system.}
	\label{fig:led}
\end{figure}

\begin{figure}[htbp] \centering
	\setlength{\abovecaptionskip}{-1pt}
	\setlength{\belowcaptionskip}{10pt}
	\includegraphics[width=6.0cm]{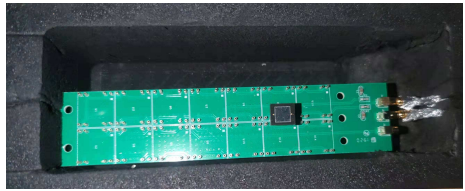}
	\caption{SiPM and support plate.}
	\label{fig:si}
\end{figure}

Measuring the radon concentration by the liquid scintillator detector of the SiPM array requires studying the pulse counts obtained on the SiPMs~\cite{f0,f2,f4}. The pulse count per unit time is proportional to the radon concentration so that the radon concentration can be determined. Also, a low-temperature measurement system was designed to study the SiPM performance characteristics (Fig.~\ref{fig:led}). In the system, two low-voltage power supplies provide the working voltage for the SiPM and the preamplifier. Also, the preamplifier amplifies the output signal by~30 times. The SiPM is placed on the support plate (Fig.~\ref{fig:si}), and they are placed together in a dark room. Then, the dark room is placed in an automatic control cryostat. The SiPM model used here is the S13360-6050CS from Hamamatsu; the working voltage is 54.86 V at room temperature, and the size is 6 mm $\times$ 6 mm $\times$ 6 mm. Its effective photosensitive area is 26.64 $\rm mm^2$, and its spectral response range is $270$–$900$ nm. 

When measuring the SPE spectrum, a pulse generator (pulser) drives the light source to illuminate the photocathode of the SiPM and outputs a pulse signal from the anode, which is driven at a frequency of $1$ kHz. Among them, a blue-emitting LED ($\lambda$ = 420 nm) is used as the light source coupled with the pulser. The pulse signal is used as the measurement signal after the fan-in fan-out. Using the synchronous triggering method, the pulser simultaneously outputs a trigger signal that passes through the low-threshold discriminator (LT-Dis). The model of this LT-Dis is N840 from CAEN. Then, a FlashADC (FADC, DT5751) from CAEN was used to sample the waveforms~\cite{dt5751}. 
%When measuring dark noise, the SiPM does not require a light source and only needs to provide a self-triggering signal to the FADC.

%During the experimental preparation, we measured the baseline of the SiPM support plate to be $2$ mV and the breakdown voltage of the SiPM to be $52$ V at room temperature. 
During the experimental preparation, we measured the preamplified baseline signal from the support plate where no SiPM was placed, which was 2mv displayed on the oscilloscope. 
We measure the breakdown voltage of the first clear signal of the SiPM at room temperature, that is, the minimum voltage at which the SiPM can operate normally at room temperature. 
Also, to reduce electrical interference and noise from the LED, the LED and its circuits are wrapped with shielding materials, such as tin foil, and fixed near the SiPM surface. The measurement of the SPE spectrum requires the pulse waveform to have a stable single photoelectron signal. It is defined as the occurrence of a single photoelectron signal once in ten pulses, $i.e.$, the probability of a single photon is $10$\%. 

In the experiments, we measured the SPE spectra with the LED through an automated cryostat to control the temperature of the experimental environment. 
%In the experiments, we measured the SPE spectra with the LED by controlling the temperature of the cryostat. 
During the data taking, the cryostat temperature and the SiPM voltage vary from $-60^{\circ}$C to $-20^{\circ}$C and 51 V to 56 V, respectively. This experiment mainly provides some reference data for the performance characteristics of SiPMs at low temperatures for developing low-background detectors for measuring the concentration of radon and its daughters and selecting the best working conditions for SiPMs to obtain high detection sensitivity. 
%Furthermore, by analyzing the amplitude and charge spectra of the SPE, we further measured parameters, such as optical crosstalk, after-pulse, gain, and SPE resolution,.

\vspace{-0.0cm}
\vspace{-0.0cm}

\vspace{-0.0cm}
\vspace{-0.0cm}

\subsection{SPE spectrum of SiPM}
\begin{figure}[htbp] \centering
%\hspace*{-29pt}
	\setlength{\abovecaptionskip}{-1pt}
	\setlength{\belowcaptionskip}{10pt}
	\includegraphics[width=7.0cm]{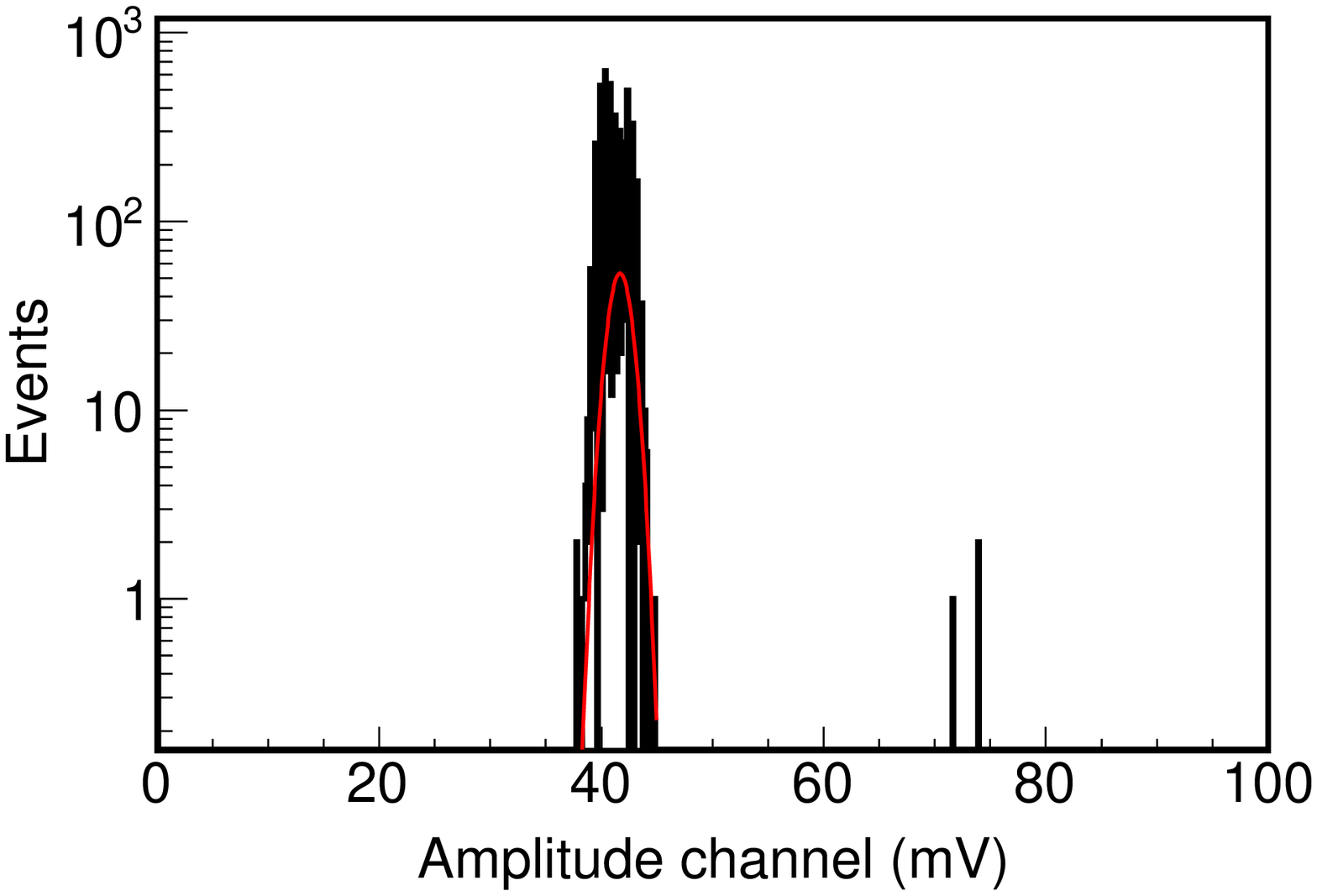}
        \includegraphics[width=7.0cm]{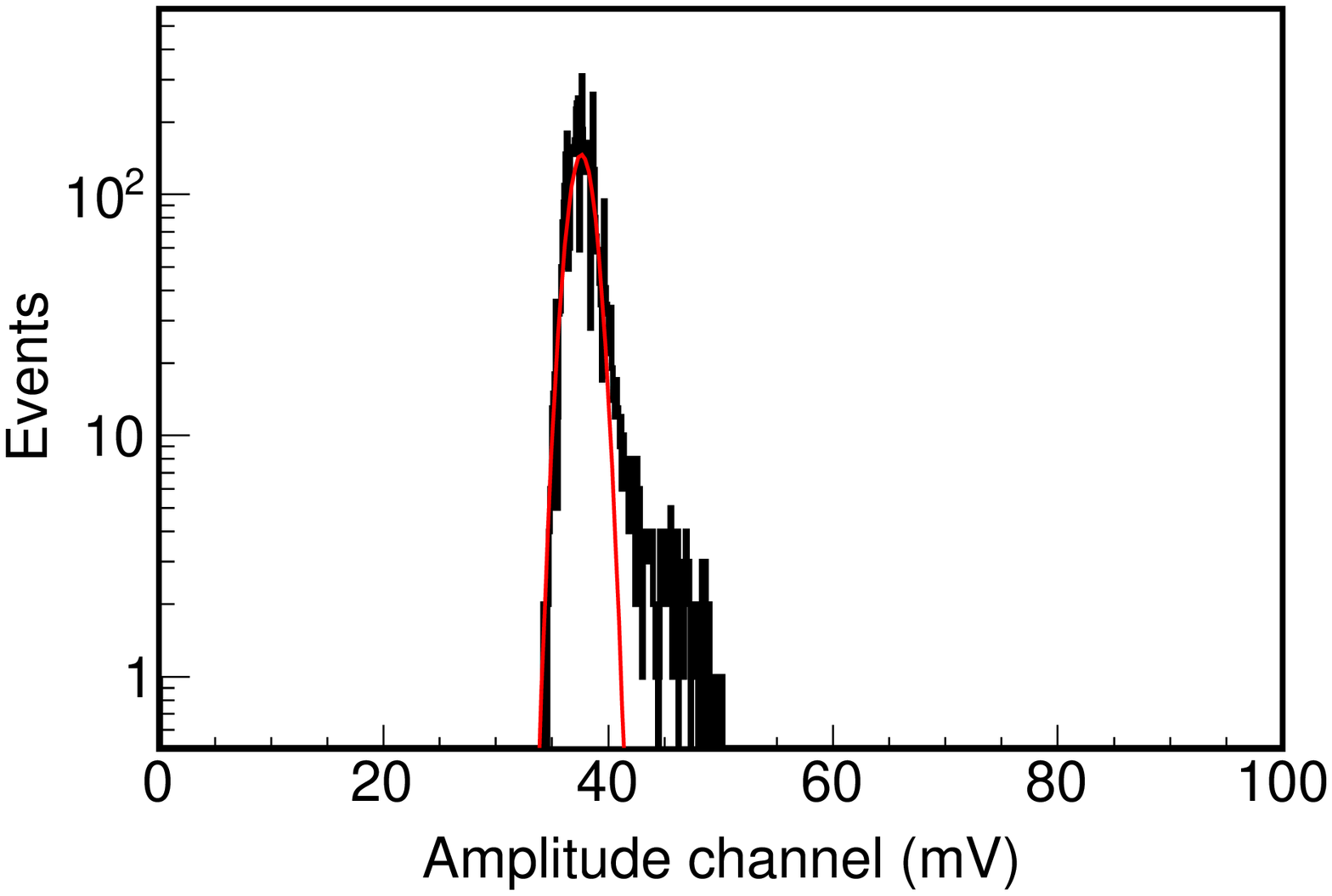}
	\caption{Amplitude spectrum with oscilloscope (left) and FADC (right).}
	\label{fig:fa}
\end{figure}

\begin{figure}[htbp] \centering
	\setlength{\abovecaptionskip}{-1pt}
	\setlength{\belowcaptionskip}{10pt}
	\includegraphics[width=7.0cm]{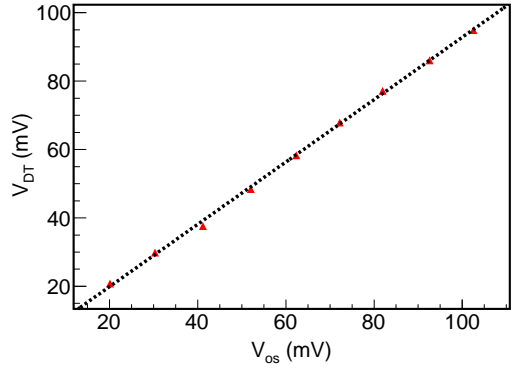}
	\caption{ FADC calibration result using amplitude spectra. Here $V_{\rm os}$ and $V_{\rm DT}$ are the mean values obtained after fitting the amplitude spectra collected by the oscilloscope and the FADC with a Gaussian function, respectively.}
	\label{fig:fadc}
\end{figure}

The SiPM waveforms are sampled by the FADC waveform automatic acquisition system, which is then calibrated. During the calibration experiment, the standard input signal is set as a standard square wave sent by the pulser, which is collected by the oscilloscope through LabVIEW~\cite{labv}, and the FADC digital acquisition system collects the output signal. Next, data sampling and analysis are performed according to the amplitude from $0$–$100$ mV, with a step size of $10$ mV. Their amplitude spectra are shown in Fig.~\ref{fig:fa}, and they are fitted with a Gaussian function to obtain the mean value as the amplitude value. Figure~\ref{fig:fadc} shows the result of the FADC calibration. The associated calibration factor obtained is $0.91$. The charge conversion equation is given as follows: 
\begin{equation} \label{b}
q=\frac{V_{\rm Fixed}}{V_{\rm FADC}\times R} \times t_0, 
\end{equation} 
where $V_{\rm Fixed}$ is the standard amplitude value corresponding to each channel ($1$ mV), and $V_{\rm FADC}$ is the amplitude value of each channel after calibration by the FADC ($0.91$ mV). $R$ is the load resistance ($50$ $\Omega$), and $t_0$ is the $1$ GHz sampling rate of the FADC. Based on the FADC calibration results, Equation~\ref{b} was used to estimate the corresponding charge value for each channel to be $21.98$ fC to analyze the charge spectrum for subsequent SPE spectrum measurements.

\vspace{-0.0cm}
\vspace{-0.0cm}

\vspace{-0.0cm}
\vspace{-0.0cm}

\begin{figure}[htbp] \centering
	\setlength{\abovecaptionskip}{-1pt}
	\setlength{\belowcaptionskip}{10pt}
	\includegraphics[width=7cm]{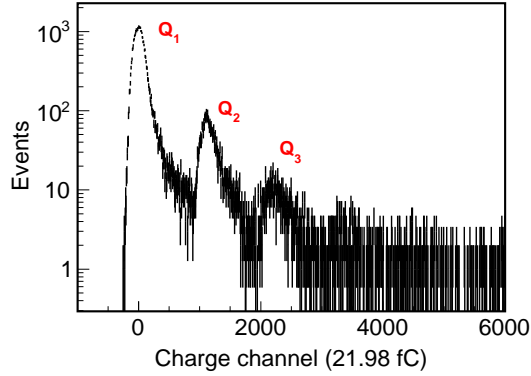}
	\caption{The charge spectra of SPE at $-20^{\circ}$C.
	}
	\label{fig:fit}
\end{figure}

\begin{figure}[htbp] \centering
	\setlength{\abovecaptionskip}{-1pt}
	\setlength{\belowcaptionskip}{10pt}
	\includegraphics[width=7.0cm]{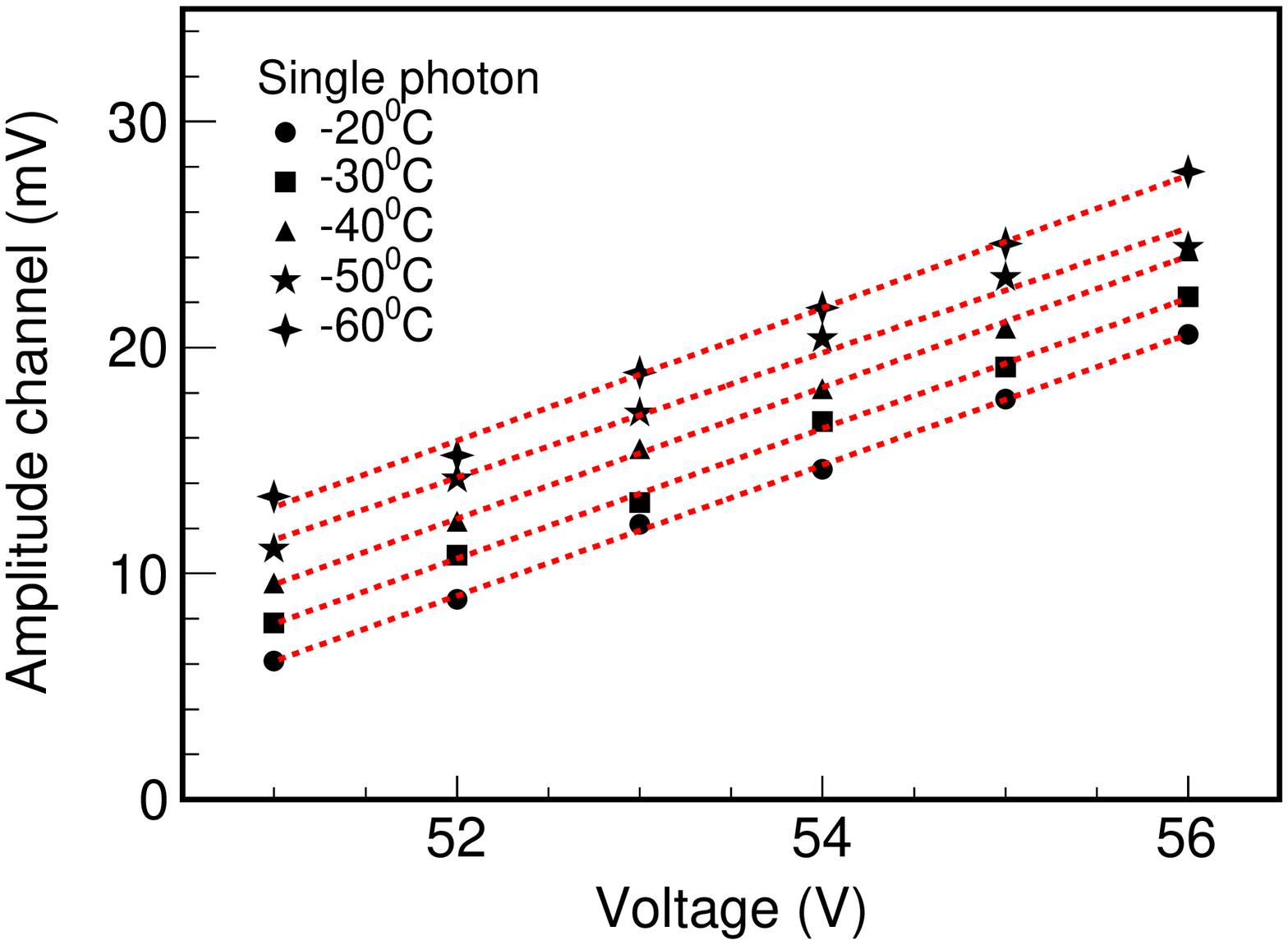}
	\includegraphics[width=7.0cm]{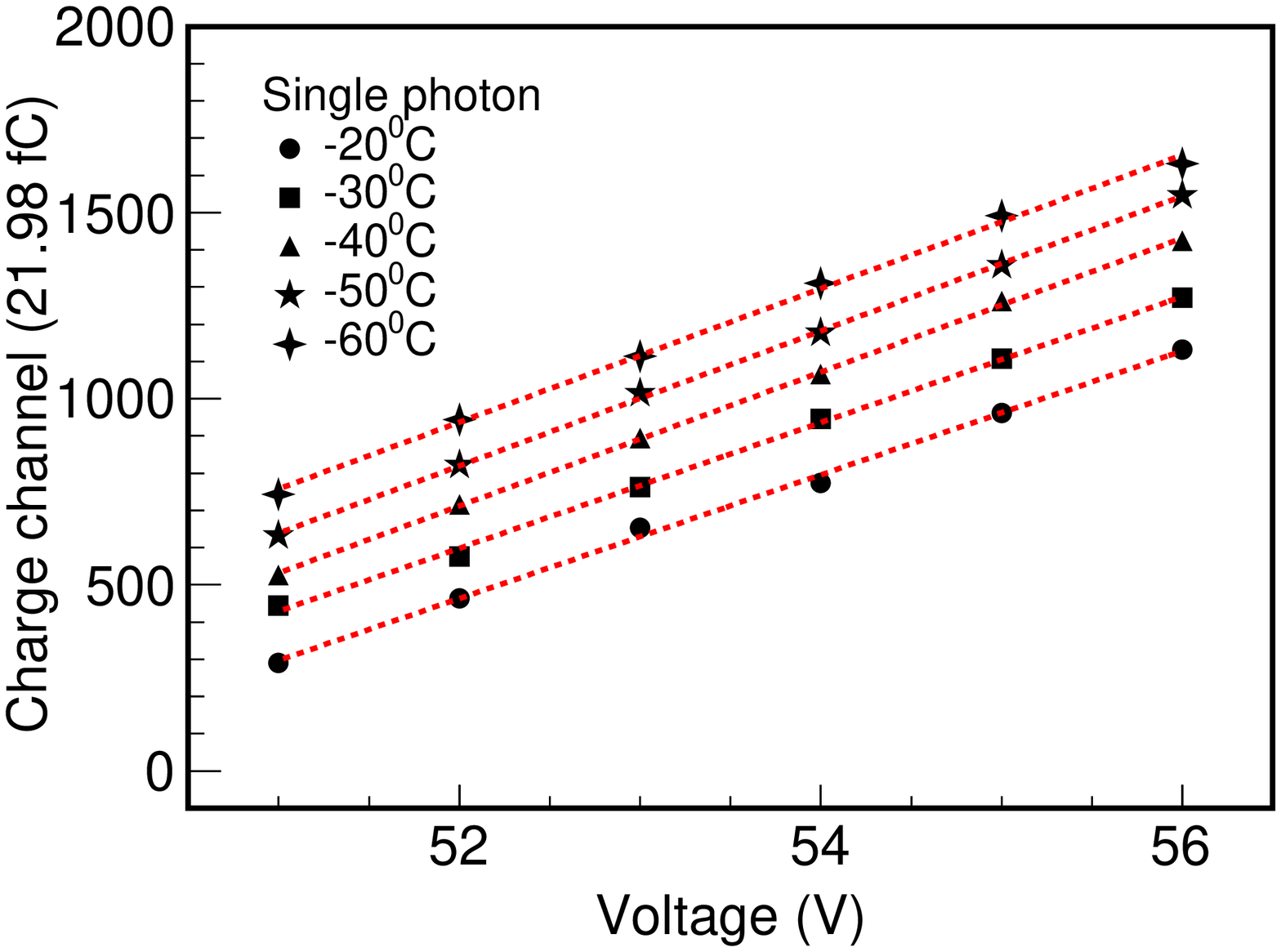}
	\caption{Amplitude (left) and charge (right) $vs.$ SiPM operating voltage distribution at different temperatures. They were fitted separately with polynomials at different temperatures.}
	\label{fig:amp}
\end{figure}

The photoelectrons generated by the photon incident on the SiPM photocathode in the detector conform to the Poisson distribution after photomultiplication~\cite{fzw}, 
\begin{equation} \label{}
P(n)=\mu^ne^{-\mu}/n! , 
\end{equation} 
where $P(n)$ is the probability that the pulse collected on the SiPM contains $n$ photons, $\mu$ is the average photon number of the pulse, and $n$ is the number of photons in the pulse. The light intensity of the LED was adjusted to ensure that $90$\% and $10$\% of the probability are a step and a signal, respectively. At this time, the ratio of the single photoelectron to the multiphoton in the signal is 1:21, $i.e.$, 
\begin{equation} \label{}
P(n=1)/p(n>1)=21 .
\end{equation} 
Also, the probability of a single photon is about $9.5$\%. Moreover, when the pulse count is large, the Poisson distribution can be approximated to a Gaussian distribution, so a Gaussian 
function can be used to fit the data. Fig.~\ref{fig:fit} shows the charge distribution of the SPE based on log coordinates, which eliminates the effect of baseline deviation and can be fitted with a multiGaussian function to obtain the peak (mean) and width (sigma) of the pulse waveform. Among them, $Q_1$ is a step mainly derived from electronic circuits and noise, $Q_2$ is a single photoelectron signal, and $Q_3$ is a two photoelectron signal. The ratio of the single photoelectron signal to the total signal can be estimated as 
\begin{equation} \label{}
 R=\frac{N_{sig}}{(N_{sig}+N_{bkg})} =9.1\%,
\end{equation} 
where $N_{sig}$ and $N_{bkg}$ are the counts of steps and signals, respectively. The results show that the signal measured in the experiment is basically the SPE, and a small amount of multiphoton signal is mixed.
The amplitude and charge spectrum of the SPE at different temperatures are analyzed in Fig.~\ref{fig:amp}, where the charge spectrum can be obtained by a simple superposition of the integral areas. Also, for the rigor of the experiment, we also examined the relationship between steps, single photoelectrons, and two photoelectrons for all acquired pulse waveforms. 
\begin{equation} \label{qqq}
Q =Q_{2}-Q_{1}, 
\end{equation}
with 
\begin{equation} \label{qq}
Q_{3} =2\cdot Q+ Q_{1} .
\end{equation}
Here, $Q$ is the channel of the single photoelectron charge. When the channels satisfy this equation, one can be sure that the data samples collected in the experiment are reasonable. Using the calibrated SPE spectra, we further investigated the performance parameters of the SiPM at different temperatures, such as gain, optical crosstalk, after-pulse, time resolution, and SPE resolution,.

\vspace{-0.0cm}
\vspace{-0.0cm}

\vspace{-0.0cm}
\vspace{-0.0cm}

\section{Results and discussion}

\subsection{Gain}
\begin{figure}[htbp] \centering
	\setlength{\abovecaptionskip}{-1pt}
	\setlength{\belowcaptionskip}{10pt}
	\includegraphics[width=7cm]{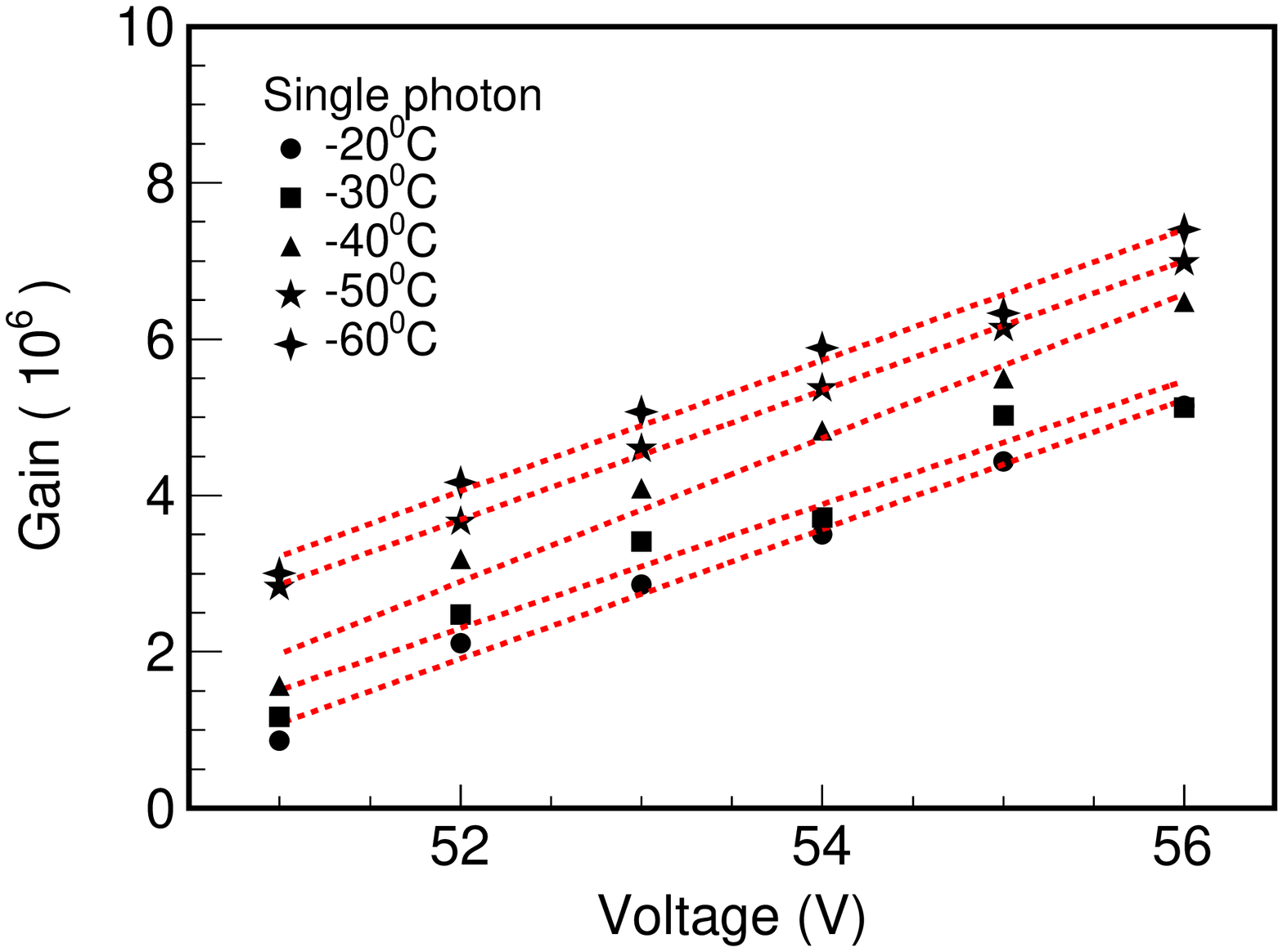}
		\includegraphics[width=7cm]{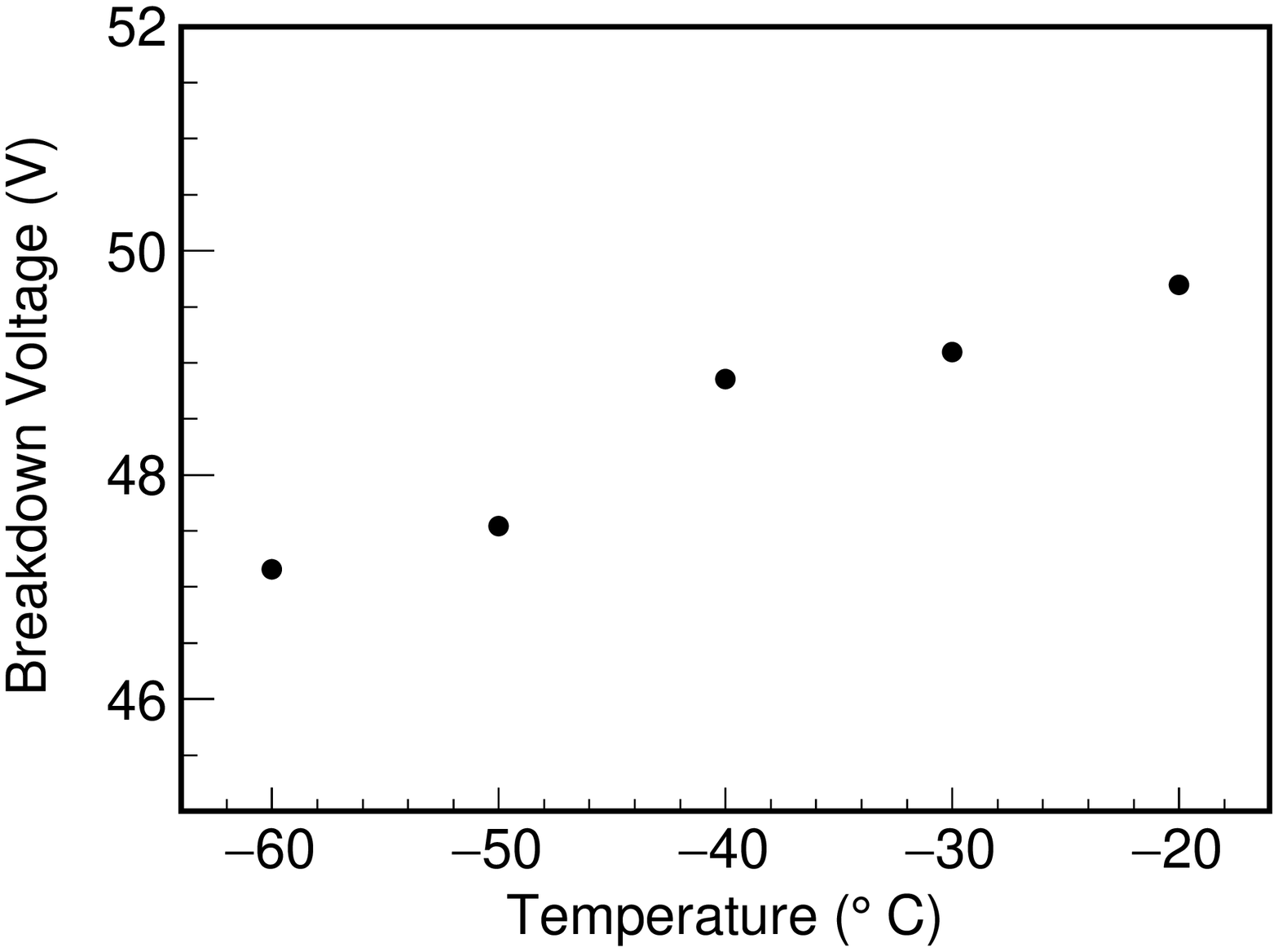}
	\caption{Gain dependence $vs.$ operating voltage of SiPM at different temperatures (left), and breakdown voltage $vs.$ temperatures (right).}
	\label{fig:gain}
\end{figure}

We evaluated the gain by illuminating the SiPM with low photon flux. SPE spectra with clearly distinguishable peaks were obtained from the SiPM anode. The gain of the SiPM is defined as follows: 
\begin{equation} \label{}
Gain=\frac{(q_2-q_1)}{e\times A} ,  
\end{equation} 
where $q_2$ is the charge of the single photoelectron ($q_2$ = $Q_2$ $\times$ $q$), $q_1$ is the charge of the electronic noise, $e$ is the charge of a single electron, and $A$ is the magnification of the preamplifier. Through the low-temperature measurement system, the temperature of the cryostat was controlled to change the working conditions of the SiPM to study the relationship between the gain and the working voltage of the SiPM at different temperatures and analyze the temperature effect on the SiIPM gain. Figure~\ref{fig:gain} (left) shows the variation of the SiPM gain with voltage under different temperatures; an obvious linear relationship exists between them. Also, as the temperature decreases, the gain increases accordingly.

\begin{figure}[htbp] \centering
	\setlength{\abovecaptionskip}{-1pt}
	\setlength{\belowcaptionskip}{10pt}
	\includegraphics[width=7cm]{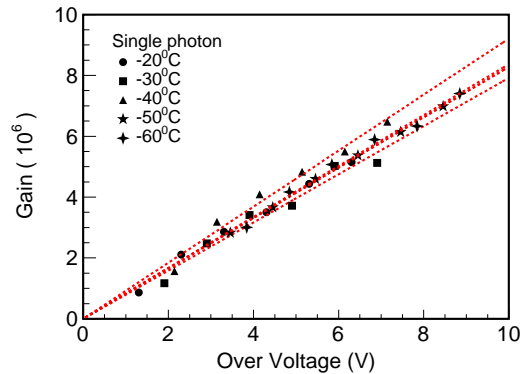}
	\caption{Gain dependence $vs.$ over-voltage of SiPM at different temperatures.}
	\label{fig:ga}
\end{figure}

%The over-voltage is the actual physical parameter that characterizes the 
%SiPM performance, which is estimated as the difference between bias voltage and breakdown voltage. 
Over-voltage is defined as the difference between the SiPM operating voltage and the breakdown voltage.
According to the gain analysis results, 
the breakdown voltage of the SiPM at different temperatures can be estimated by extending the fitted line shape outwards in the Fig.~\ref{fig:gain} (left), and calculating their corresponding over-voltages in the Fig.~\ref{fig:ga}. Figure~\ref{fig:gain} (right) shows the variation trend of breakdown voltage with temperatures. It can be found that the decrease of temperature can reduce the corresponding breakdown voltage. 
\vspace{-0.0cm}
\vspace{-0.0cm}

\vspace{-0.0cm}
\vspace{-0.0cm}

\subsection{Optical crosstalk}
During the avalanche process, the photons emitted by the carrier have a certain probability to reach the adjacent cell diode and trigger a second avalanche. The probability of an avalanche caused by this mechanism is called optical crosstalk probability~\cite{afop1,sipm2}. 
%During the avalanche process, a small number of carriers that can be absorbed by the Si material are absorbed by the adjacent cells, resulting in the probability of a secondary avalanche, $i.e.$, optical crosstalk~\cite{afop1,sipm2}. 
The optical crosstalk limits the photon counting resolution of SiPMs. Therefore, the optical crosstalk probability is an important SiPM property that should be reduced as much as possible. Methods for determining optical crosstalk are based on the SPE spectrum analysis. We estimate the optical crosstalk probability as,  
\begin{equation} \label{}
P_{ct}=N_{1.5 p.e.} /N_{0.5 p.e.} ,  
\end{equation} 
where $N_{1.5 p.e.}$ is the count of optical crosstalk in all single photoelectron pulses, $i.e.$ above the $1.5$ photoelectron ($p.e.$) threshold, and $N_{0.5 p.e.}$ is the count of all pulses after step removal, $i.e.$ above $0.5 p.e.$ threshold.
%$0.5$ photoelectron, which is the total pulse signals after removing the steps. 
%the count of $1.5$ photoelectron ($p.e.$), and $N_{0.5 p.e.}$ is the total count of all pulses. 

\begin{figure}[htbp] \centering
	\setlength{\abovecaptionskip}{-1pt}
	\setlength{\belowcaptionskip}{10pt}
	\includegraphics[width=8cm]{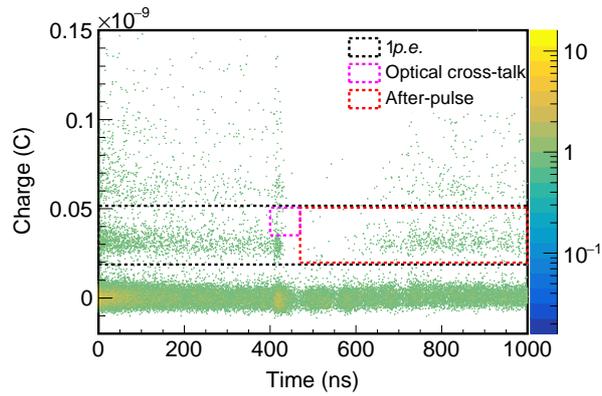}
	\caption{The charge $vs.$ time corresponding to the maximum peak at $-40^{\circ}$C. 
	Here, the black dashed box is the signal of the SPE, 1$p.e.$, while the red box is the after-pulse signal in the SPE and the pink box is the optical crosstalk signal in the SPE.
	%Here the the pink and red boxes mark the optical crosstalk and after-pulse events.
	}
	\label{fig:scatter}
\end{figure}

\begin{figure}[htbp] \centering
	\setlength{\abovecaptionskip}{-1pt}
	\setlength{\belowcaptionskip}{10pt}
	\includegraphics[width=7.0cm]{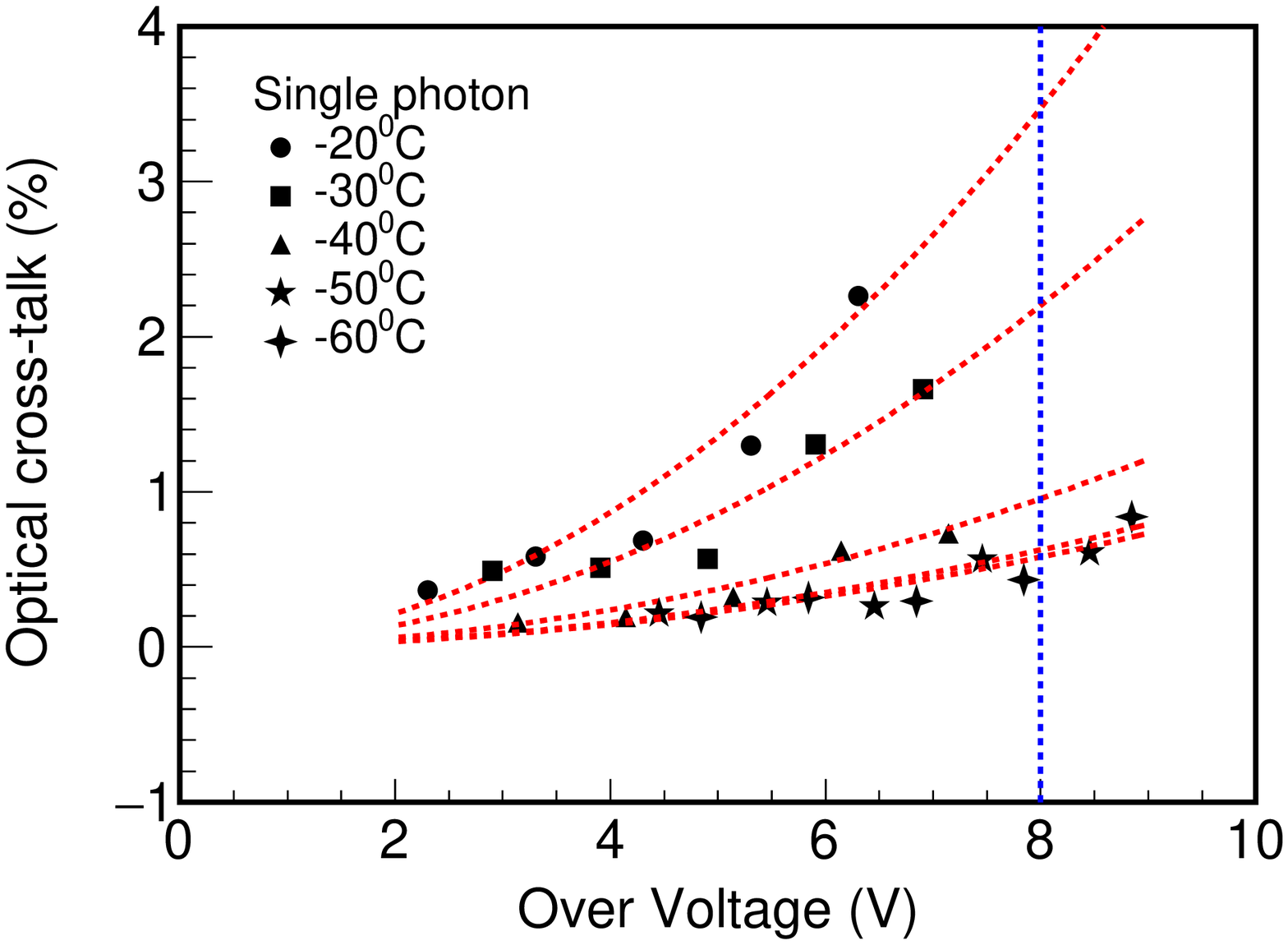}
		\includegraphics[width=7.0cm]{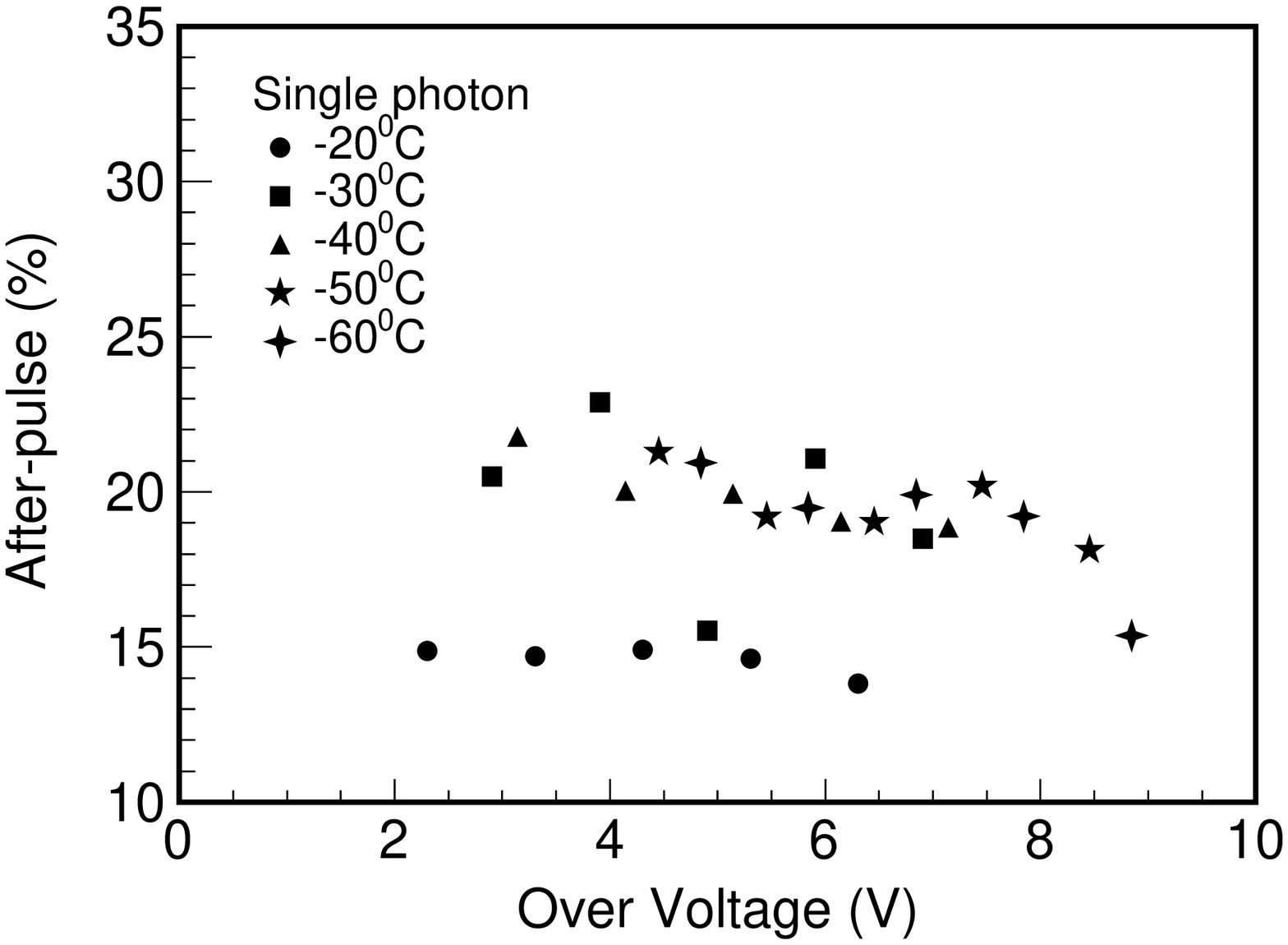}
	\caption{Optical crosstalk (left) and after-pulse (right) at different temperatures.}
	\label{fig:op}
\end{figure}

\begin{figure}[htbp] \centering
	\setlength{\abovecaptionskip}{-1pt}
	\setlength{\belowcaptionskip}{10pt}
	\includegraphics[width=7.0cm]{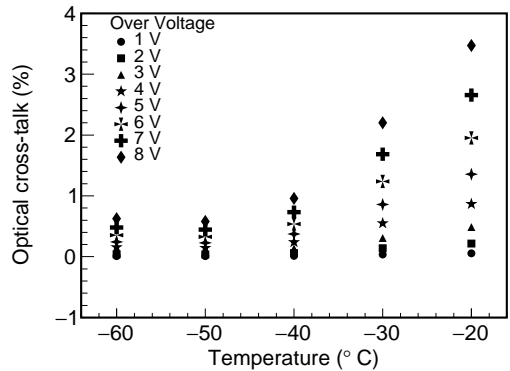}
		\caption{Optical crosstalk $vs.$ temperature at different over-voltage.}
	\label{fig:ct}
\end{figure}

In the experimental analysis, it is usually difficult to distinguish the optical crosstalk and the after-pulsing. A two-dimensional histogram of the charge and time corresponding to the maximum peak in the pulse waveform were used to study them (Fig.~\ref{fig:scatter}). Evidently, a single photoelectron after-pulse signal and a small amount of optical crosstalk signal are observed. To estimate the counts of the optical crosstalk above the $1.5$ $p.e.$ threshold, the time must be within the pulse signal range (between 400 and 470 ns), and the charge must be a smear to the right of the single photoelectron peak. We calculated the optical crosstalk rate of a single photon at different temperatures (Fig.~\ref{fig:op}) and found that although the optical crosstalk probability is very low. 
It has a clear law of increasing change with the increase of temperature in Fig.~\ref{fig:op} (left) and Fig.~\ref{fig:ct}. 
Especially when the temperature is from $-20^{\circ}$C to $-40^{\circ}$C, this change is very clear, that is, as the temperature decreases, the optical crosstalk probability also decreases accordingly. 
%As the temperature decreases, the optical crosstalk rate decreases accordingly. 
However, the probability of optical crosstalk at temperatures below $-40^{\circ}$C is less affected by temperature. That is, the optical crosstalk tends to saturate under the influence of low temperature conditions.  
Therefore, the low-temperature environment can greatly reduce the optical crosstalk interference of SiPMs. 
Moreover, when the over-voltage is lower, the probability of optical crosstalk is relatively lower, especially when the over-voltage is lower than 3V.

\vspace{-0.0cm}
\vspace{-0.0cm}

\vspace{-0.0cm}
\vspace{-0.0cm}

\vspace{-0.0cm}
\vspace{-0.0cm}

\subsection{After-pulse}
The measurement of the after-pulse rate is also based on the SPE spectrum analysis. After-pulses are electrons generated during the avalanche that are captured and released again after a delay lasting from nanoseconds to microseconds, resulting in a new secondary current pulse with a smaller amplitude than the original secondary current, $i.e.$, 
%the pulse amplitude will be less than 1 $p.e.$ " small pulse".
a “small pulse” whose the pulse amplitude will be less than the real signal~\cite{afop1,sipm2,afop2}. 
Unfortunately, the after-pulse signals cannot be separated from the true single photoelectron signal, reducing the photon counting resolution. Therefore, the effect of different temperatures on the after-pulse rate must be studied to minimize its interference with the real signal. 

Fig.~\ref{fig:scatter} clearly distinguishes the step (0$p.e.$), single photoelectron (1$p.e.$), and two photoelectron signals (2$p.e.$). To estimate the after-pulse count of a single photoelectron, the time must exceed 470 ns, and the charge must be within the range of the channel corresponding to the valleys on both sides of the single photoelectron peak, $i.e.$ between 0.5$p.e.$ and 1.5$p.e.$. By estimating the fraction of after-pulse counts in 0.5$p.e.$ photoelectron counts, we obtained the after-pulse rates at different temperatures in Fig.~\ref{fig:op} (right). 
 From the behaviour observed, the after-pulse
probability has no dependence on the temperature change when the temperature is between $-60^{\circ}$C and $-20^{\circ}$C.

\subsection{SPE resolution}
\begin{figure}[htbp] \centering
	\setlength{\abovecaptionskip}{-1pt}
	\setlength{\belowcaptionskip}{10pt}
	\includegraphics[width=7cm]{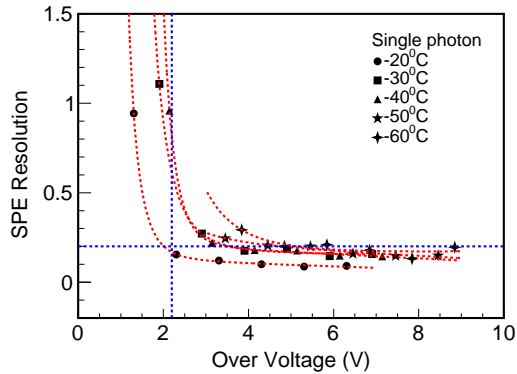}
	\caption{SPE resolution distribution at different temperatures.}
	\label{fig:re}
\end{figure}

When the single photoelectron signal is measured, most of the signal obtained by the FADC is the electronic noise, which overlaps with the single photoelectron signal and even other multiphoton signals. 
%Therefore, other signals will be unavoidable when acquiring a single photoelectron signal. 
As many photons are detected on the SiPM, the mean value of the single photoelectron peak obtained by fitting the estimated charge distribution deviates from the mean value of the real single photoelectron distribution, thus affecting the SPE resolution of the charge measurement. The SPE resolution of a single photoelectron is defined as 
\begin{equation} \label{}
\delta=\frac{\sigma_{\rm \tiny SPE}}{mean_{\rm \tiny SPE}}, 
%\delta=\frac{S}{\Delta M}, 
\end{equation} 
where $\sigma_{\rm \tiny SPE}$ is the sigma value after fitting the SPE charge spectrum. And $mean_{\rm \tiny SPE}$ is the mean value after fitting the SPE charge spectrum after deducting steps. The measured SPE resolution results at different temperatures are shown in Fig.~\ref{fig:re}. 
The result analysis revealed that the SPE resolution is affected by temperature. 
It can be clearly found that when the over-voltage is between 0V and 3V, SPE resolution is improved rapidly with the increasing of over-voltage. However, when the over-voltage is between 3V and 8V, the SPE resolution is slightly improved with the increasing of over-voltage.
%It can be clearly found that the SPE resolution is better with the decrease of temperature when the over-voltage is  between 0V and 3V. 
%However, when the over-voltage is between 3V and 8V, the SPE resolution gradually improves with the increase of over-voltage, and it is better than 0.2.
%However, the SPE resolution tends to be stable with the increase of over-voltage, and is basically around 0.2 when the over-voltage is greater than 3V. 
%But as the over-voltage increases, the SPE resolution gradually remains stable. When the over-voltage is greater than 3V, it basically saturates to about 0.2. 
Therefore, the SPE resolution of SiPM can be improved by reasonably controlling the temperature and the over-voltage. 

\vspace{-0.0cm}
\vspace{-0.0cm}

\vspace{-0.0cm}
\begin{figure}[htbp] \centering
	\setlength{\abovecaptionskip}{-1pt}
	\setlength{\belowcaptionskip}{10pt}
	\includegraphics[width=7cm]{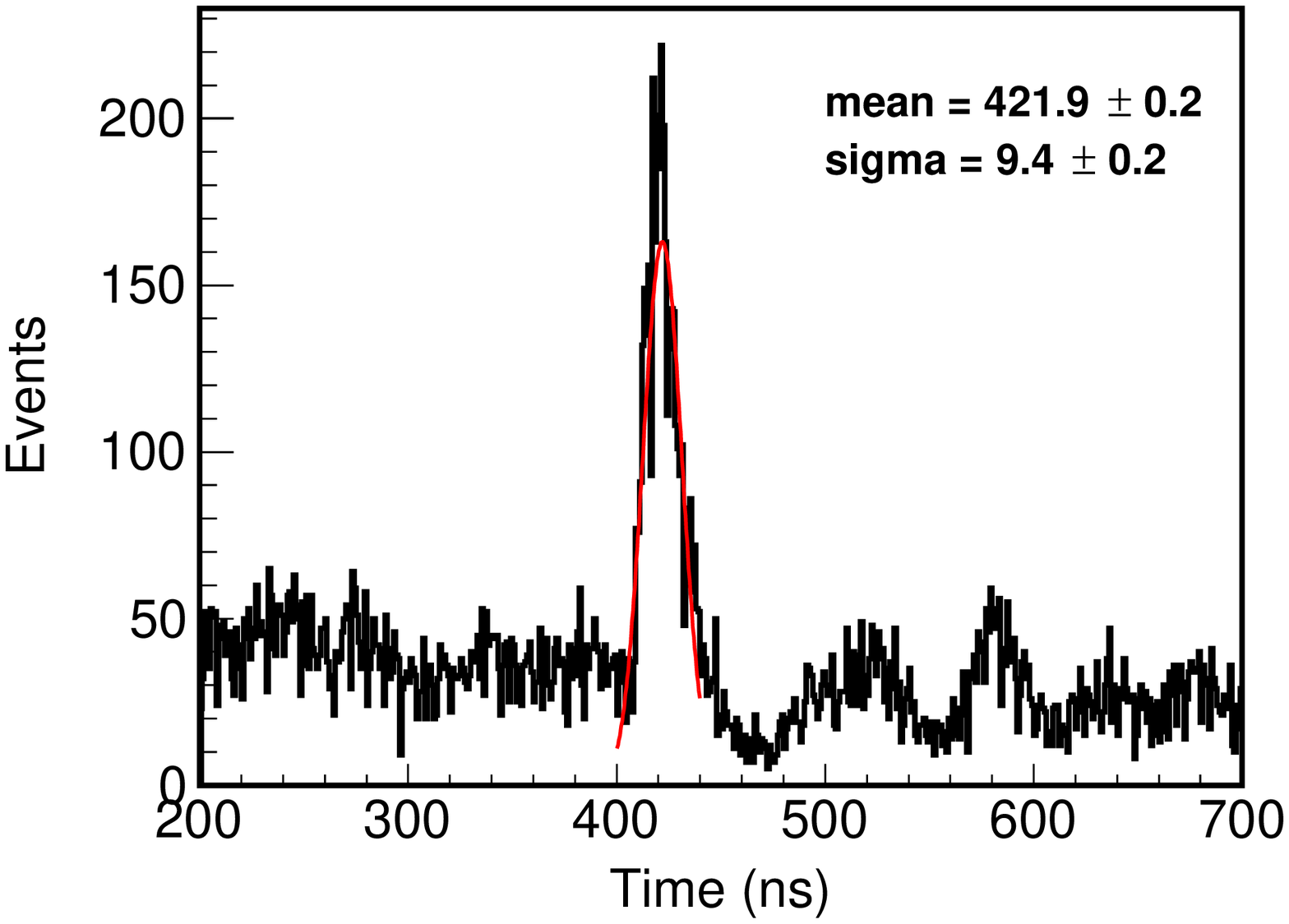}
		\includegraphics[width=7cm]{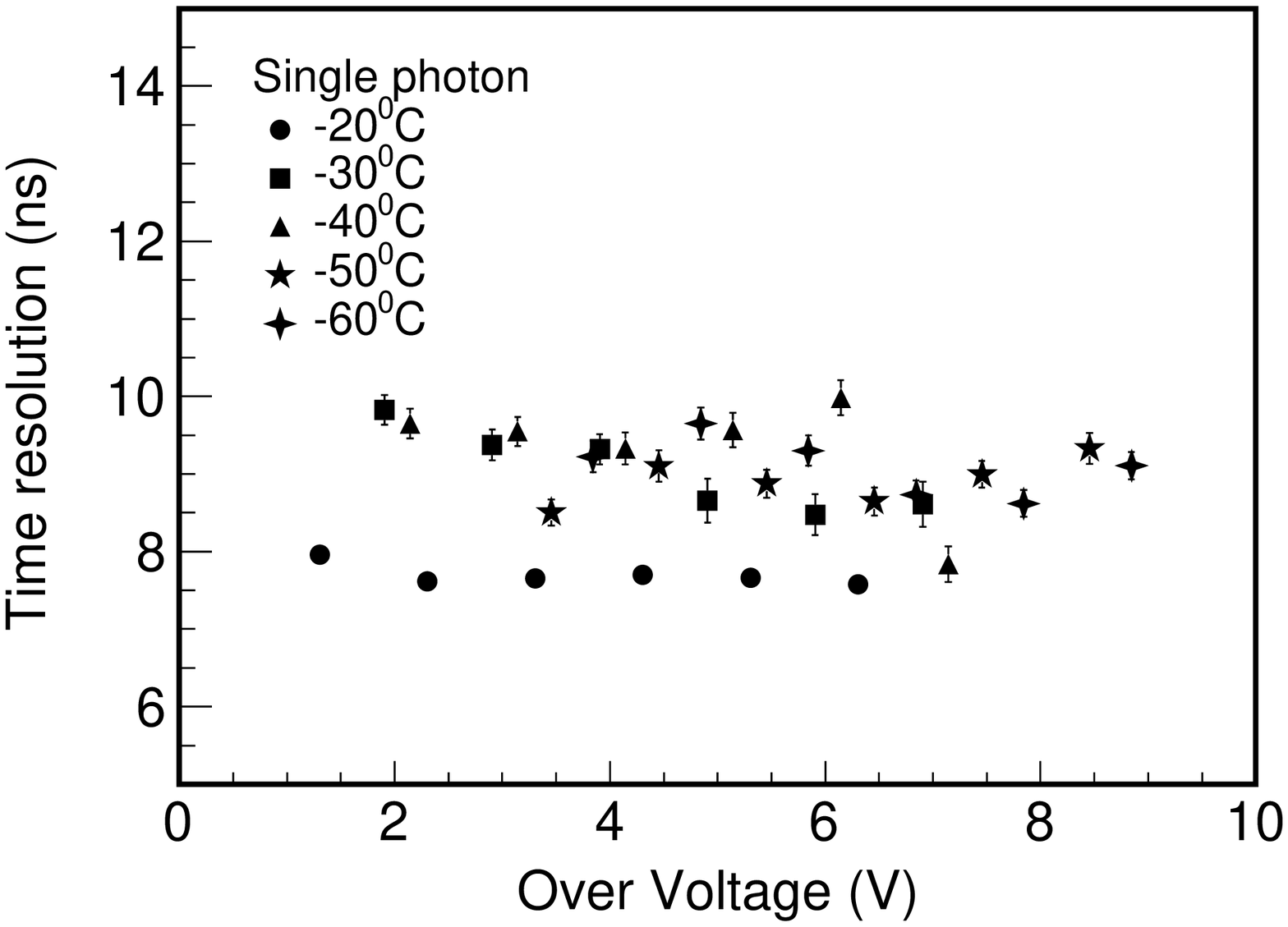}
	\caption{Time distribution of the SPE at 51V with $-20^{\circ}$C (left), and the time resolution at different temperatures (right).
%	Time distribution at $-20^{\circ}$C (left), and the time distribution at different temperatures (right).
	}
	\label{fig:ti}
\end{figure}
\vspace{-0.0cm}

\subsection{Time resolution}
SiPMs are photodetection devices with fast time responses. In the experiments, the transit time and transit time dispersion, which characterize the temporal characteristics, were studied. They refer to the time from the light incident event on the photocathode surface to the appearance of the output pulse and fluctuations in the transit time of all single photoelectron pulses on the photocathode surface, respectively. 

During the pulse waveform analysis, the transit time was challenging to estimate because the baseline shifted generally. 
Therefore, the time resolution was used to study the time characteristics of the SiPM in the pulse analysis. 
In Fig.~\ref{fig:ti} (left), it is the time distribution corresponding to the maximum amplitude value of each pulse waveform. However, time resolution is defined as the sigma value obtained by fitting the time distribution with a Gaussian function. 
In this paper, the time resolution depends only on the readout scheme without considering the effect of the electronics on the SiPM alone. 
%The time resolution is defined as the sigma value obtained by fitting the time resolution corresponding to the maximum pulse peak value in the pulse waveform with a Gaussian function (Fig.~\ref{fig:ti} (left)). 
%Therefore, the time distribution to study the time characteristics of the SiPMs in the pulse analysis is the sigma value obtained by fitting the time distribution corresponding to the maximum pulse peak value in the pulse waveform with a Gaussian function (Fig.~\ref{fig:ti} (left)). 
In the figure, the photon signal is mainly concentrated between the time 400–480 ns, and the SiPM time resolution of SiPM is $(9.4\pm0.2)$ ns. By studying the time resolution of different temperatures, it was observed that the time parameter is not affected by temperature (Fig.~\ref{fig:ti} (right)).

\section{Conclusion}
In conclusion, we study how to realize the high sensitivity of liquid scintillation detector for very low concentration radon measurement by reducing the influence of SiPM noise and improving the light yield of liquid scintillation. 
In order to obtain the optimal working conditions of SiPM to improve its parameter performance and reduce the impact of its noise, we built an automatic and accurate low-temperature measurement system to study the SPE spectrum, SPE resolution, optical crosstalk, and after-pulse of SiPM at different temperatures. 

In the experiment, we found that the low temperature environment can effectively improve the performance of SiPM. When the temperature is lower than $-40^{\circ}$C and the over-voltage is less than 3V, the optical crosstalk probability of SiPM is low, and the SPE resolution is relatively better. Therefore, we take it as the optimal condition, and can reduce the noise of SiPM by reasonably controlling it. 
Moreover, we can also know that there is a strong dependence between temperature and the light yield of liquid scintillation. When the experimental temperature is lower than room temperature, the light yield of liquid scintillation will increase accordingly. Therefore, when the temperature is lower than $-40^{\circ}$C, we can obtain a higher light yield, thus improving the liquid scintillation sensitivity. 
By rationally utilizing this optimal working condition, we can effectively reduce the noise of SiPM, improve the light yield of liquid scintillation, and thus improve the sensitivity of the detector and meeting the experimental requirements of extremely low-background detection.

\section{Acknowledgment}
The authors wish to thank all teachers and classmates of the School of nuclear science and technology, University of South China. This work is supported by the State Key Laboratory of Particle Detection and Electronics, SKLPDE-KF-202203; Special funds for the construction of innovative provinces in Hunan under contract No. 2020RC3054; Hunan Postgraduate Research Innovation Project in 2021, CX20210914.

%\bibliographystyle{99}
%\balance

\end{document}